# Non-Parametric Detection of Network Communities; The Natural Way – A Cascaded Stackelberg Game

Nishant Deepak Keni
School of Electrical and Computer Engineering
Georgia Institute of Technology
Atlanta, Georgia, U.S.A.
nishant.keni@gatech.edu, asingbal3@gatech.edu

*Abstract*— Real-World networks have an inherently dynamic structure and are often composed of communities that are constantly changing in membership. Identifying these communities is of great importance when analyzing structural properties of networks. Hence, recent years have witnessed intense research in of solving the challenging problem of detecting such evolving communities. The mainstream approach towards community detection involves optimization of a global partition quality metric (e.g. modularity) over the network [1]. Optimization of this global quality metric is often akin to community assignment by a centralized decision-making model. Another technique, Spectral Clustering [2], involves mapping of original data points in a lower dimensional space, where the clustering properties of a graph are much more evident, and then applying standard clustering techniques such as K-Means or Fuzzy C-means for identifying communities. However, the traditional spectral clustering techniques cannot naturally learn the number of communities in networks [1]. These techniques are based on external community connectivity properties such as graph cuts, and often fail to identify smaller community structures in dense networks. In this article, we propose an algorithm, namely, the Cascaded Stackelberg Community Detection Algorithm (CASCODE) inspired by the Stackelberg Duopoly Game [3]. This algorithm uses the notion of a leader-follower relationship between the nodes to influence the actions of either. The intuition of the algorithm is based on the natural expected internal structure in evolving communities in networks. Thus, the algorithm is able to naturally learn the number of communities in a network in contrast with other techniques such as Spectral Clustering, which require the expected number of communities as an input. Because this Stackelberg Model-based Community Detection algorithm detects communities through their internal structure, we are able to obtain a finer community structure resolution in dense networks.

*Keywords— Non-Parametric Community Detection; Game Theory; Stackelberg Duopoly Game; Clique; Modularity Optimization; Normalized Mutual Information*

## I. Introduction

In the theory of network analysis, the concept of a community doesn't have a ubiquitous definition. The definition of what a community is depends on the system and the application in consideration. However, a general intuition of a community is what is constituted by a group of cohesive nodes, such that, the density of links between the nodes inside this group is more than those connecting the nodes outside this cohesive group [2]. Researchers in social network analysis approach community detection as a problem of decomposing a network into modules of nodes that best capture the network's modular tendencies. This property of modularization acts as a key to understanding a wide range of phenomena related to network structures, as these modules represent meaningful (and sometimes independent) units of a network organization. Understanding the formulation and evolution of communities in an important research topic in sociology because of its important relations to the studies of social marketing, criminology, urban development, etc. For example, for effective online marketing, such as placing ads or deploying viral marketing strategies, identifying communities in the social network leads to accurate targeting leading to better marketing [4].

Community Detection in Networks, being a long-standing research problem, several techniques exist which are sheerly based on network structures. For example, many algorithms either involve a global optimization of heuristics such as modularity, betweenness, conductance or just look for network partitions with predetermined network structures [2]. Optimization of a centralized global metrics often overrides individual node preferences and enforces the networks structural decomposition from a centralized viewpoint. Such algorithms do not support the theory underlying the systematic evolution of networks as an ensemble of several communities. Often in social networks, communities are formed bottom up, without a global authority trying to enforce a global optimal objective [1]. Some community detection algorithms, as mentioned above focus on partitioning the graph into disjoint components, thus assigning hard membership to nodes to their respective communities. This kind of membership seldom occurs in real social networks, where an individual typically belongs to multiple communities, such as its family, friends, colleagues, etc. Thus, an ideal community detection algorithm should encapsulate the intuition of overlapping communities and attempt to identify communities by identifying the natural way in which they evolved.

In this article, we approach to solve the above-mentioned issues by borrowing ideas from Game Theory, which deals which deals with the study of the strategic interaction between self-interested agents. To adopt this framework, we treat nodes as rational independently functioning entities. Using this theoretical construct, we attempt to model the evolution of a network using the Stackelberg Game [3]. The Stackelberg Game is built upon the notion of a leader and a follower, in which, the follower observes its leader's actions and chooses to act in a way so as to maximize its utility. On the other hand, the leader knows the follower's actions ex-ante and acts in order to maximize its utility. The leader-follower game can either be a co-operative or a non-co-operative one where both either work in unison to maximize their payoffs, or they act selfishly against each other to achieve their maximal. We build our community detection algorithm over the Co-operative Stackelberg Game framework where the leaders move first individually and the followers choose their leaders and follow their actions sincerely. Since this algorithmic framework is non-parametric, it does not expect any apriori in the expected number of communities and has no resolution limit with regard to the community size. Also since this algorithm uses a bottom-up framework, trying to model the formation of communities in an evolving network, it exploits a bottom-up approach to community detection, it is guaranteed to find certain community structures in the network efficiently if they exist. We describe these community structures below.

The notion of a community carrier two main intuitions with it. First, we expect that the nodes in the community form a clique, viz. show dense connectedness. Second, we expect that there exist some nodes which do not have links to nodes outsides its community, viz. there exist some nodes which are unique to their respective communities. These properties are internal to a community and are characteristics of dynamic community evolution. As mentioned before, community detection techniques such as spectral clustering are not able to exploit these properties as they function by optimizing external connectivity between communities to identify optimal partitions of the network. This is evidenced by the fact that these techniques fail to identify small communities in dense networks. Contrary, in this article we build a framework which focusses on exploiting the network community structures whose formation is motivated by above internal properties so that we can efficiently identify communities in the network through a bottom-up approach. We also test the robustness of our algorithm on generic networks with no succinct community structures or those formed without preferential attachments.

## II. Prior Research In Community Detection Algorithms

Finding communities in a network can be empirically thought of as a clustering problem, where nodes are assigned to a community (cluster centroid) in some suitable manner such as using K-Means or Fuzzy C-Means [1]. While the former imposes a hard membership on the nodes to a particular community, the latter allows nodes to possess memberships to multiple communities weighted by some importance metric. A lot of research has been done in the area of community detection which has resulted in benchmark algorithms. We describe some of the relevant ones here. The classic method for community detection is Spectral Clustering which attempts to partition data into a lower-dimensional representation obtained by projecting the network's similarity matrix on the top k eigenvectors of the graph Laplacian matrix. After embedding the original data onto a lower dimensional representation, the final clustering result leading to community assignment is obtained by running one of the above clustering algorithms on this representation [1].

Another variant of Spectral Clustering (E.g. RatioCut) exploits the concept of graph cuts to introduce partitions in a graph [1]. These algorithms operate by minimizing the inter-community edges when nodes are grouped into disjoint communities. However, minimizing graph-cuts being an NP-hard problem, Spectral Clustering is often performed approximately using the technique mentioned in the previous paragraph. However, the major weakness of Spectral Clustering algorithms for community detection is that they require apriori specification of the number of communities. This information may not be known or estimated in many cases dealing with large complex networks. Also, Spectral Clustering has a poor resolution to detect small communities in dense networks.

Many of the existing literature in Game-Theoretic Frameworks for Community Detection models each node in a network as a rational agent trying to optimize its own utility by joining or leaving communities, which it defines as a community formation game [4][5][6]. A Nash equilibrium of the game can then be readily interpreted into a community structure of the network: the communities every node belongs to in a Nash equilibrium becomes the output of our community detection algorithm. Due to computational constraints, the local equilibria are used instead of Nash equilibria as solutions. The game-theoretic framework reflects real-world organic community formations, and thus in principle is a more systematic framework than others that rely only on global optimization objectives or have no objectives. This framework also naturally incorporates overlapping communities, because it allows each individual agent to select multiple communities, as long as it could improve its utility. As a result, we are able to model node preferences quantitatively in a network to resolve finer community structures formed due to rational attachments.

## III. Methodology

Let's now begin to frame our community detection problem as a Stackelberg Game [3]. The Stackelberg Game as mentioned before is a leader-follower sequential game. Thus, to model develop a framework based on this game, we must first be able to distinguish nodes of the network as being leaders or followers. In the context of Community Detection, we can hypothesize every community to consist of a leader and other nodes in the community as being followers of this leader. We can go a step further to involve the intuition of dynamically evolving communities. We can state that a network is constructed by a merger of communities such that the leaders (we call these prime leaders as central leaders) are the point of this central connectivity. The immediate neighbors of these leaders are then its followers (we call them co-leaders), who again act as leaders to their respective neighbors. This recursion will continue until we encounter terminal nodes, who will have no followers and hence cannot be termed as co-leaders. These nodes simply act as follows to their respective co-leaders. Thus, intuitively the evolution of a multi-community dynamic network can indeed be modeled as a sequential Stackelberg Game between every leader and its follower, such that the Game propagates outwards from Epicenter of a multi-community network.

Using the above intuition as a base to our framework, it is important to realize that game starts from those leader nodes which are the point of central connectivity in the network. We can use this definition recursively to identify co-leaders within the communities led by the prime leaders. However, note that only those partitions associated with these central leaders are desired communities. The notion of sub-leaders and the sub-communities associated with them are only to find the solution to our problem by iterative application of the Stackelberg Model to the interaction between the nodes to form communities. We also mentioned before that any intuitive evolving dynamic community would have a node that is connected to only nodes lying in the community to which it belongs to. Using the notion of co-leaders and the sub-communities associated with them as we go identifying communities bottom-up, we would come across nodes which would only be linked to other nodes in its sub-community. At every application of our Stackelberg Model, we call such nodes as followers. Thus, if we are able to find a membership to qualify a node as a leader or a follower then we will be able to identify the communities that make up the network.

The notion of followers as defined above states that followers are only connected to nodes in their respective sub-communities. Thus, to have a connection (interaction, in game theoretic terms) they would have to take resort of the links which go through their co-leaders and ultimately the central leaders. This inference helps us to use betweenness centrality to provide memberships to leaders. Thus, intuitively, we expect that the central leaders will have the highest betweenness centrality, followed by the co-leaders (which are actually followers of their respective leaders, terminating ultimately at the central leaders) and followed by the followers. This idea now gives a firm footing to identify leaders and followers.

We begin by calculating the betweenness centrality of each node in the network. We then define leaders as nodes which have the highest betweenness centrality amongst all its neighbors. The remaining nodes are then followers associated with their respective (neighboring) leaders. Once the leaders and the followers have been detected, we now perform community assignment to them. To do so, we sort the leaders in decreasing order of their betweenness centrality. Using the ideas introduced before, we are sure at this stage, that the top-most leaders in the sorted list are of course the central leaders. We then assign a new community membership label to all the leaders, such that the first leader in the sorted list (having the highest betweenness centrality) is assigned a community label 1, followed by another leader, who hosts community 2 and so on. Thus, we have a new community assigned to all these leaders. We now seed the communities with followers, headed by their respective co-leaders. This is done by assigning the same respective community membership label to the neighbors of these leaders, which are actually its followers. Concretely, neighbors of the Leader 1 Node, (who heads community marked by label 1) will all be assigned to community label 1 and so on. These followers then act as co-leaders to their followers (neighbors, who are not their leaders or are not assigned any membership yet, although the 2 cases would often overlap) and assign the same respective memberships to them. By continuing so for all the nodes in the network, we end up giving a community membership to all the nodes in the network.

We have now performed a run of our community detection algorithm assigning appropriate community memberships to the nodes. However, in a general network, we may encounter a case where a co-leader doesn't have any associated follower. In such a case, we assign it the community membership (as a follower of that co-leader) that a majority of its neighboring nodes are associated with ties broken randomly. Indeed, if a network satisfies the structural properties of having community cliques with at least one node having a fully internal connectivity, we are guaranteed to assign an accurate community membership to all the nodes in the network. We prove this claim in a later section.

We now formalize the proposed algorithm in succinct steps. Consider a network graph G (V, E), with node set V and edge set E, such that a community C (a subset of G) is a community such that C is a clique and has at least one node which has fully internal connectivity, viz. has no links to any node outside C (belonging to C'). Mathematically, this community structure can be represented as $C = \{n: n \subseteq V$ and $(n_i, n_j) \subseteq E \ \forall n_k \subseteq C$ and $\exists \ i$ such that $n_j \subseteq C \ \forall \ (n_i, n_j) \subseteq E\}$. The Cascaded Stackelberg Game as runs as follows –

*1. Compute the betweenness centrality of all the nodes in the network.*

*2. Form a leader set L such that every leader $l \subseteq L$, has the highest betweenness centrality amongst its neighbors, with ties broken randomly. The remaining nodes form the follower set as F = V\L.*

*3. Assign a unique community label to every leader node $l \subseteq L$.*

*4. For every node $l \subseteq L$, assign the same community membership to its neighbors N, where $N \subseteq F$. These nodes act as co-leaders and can now be a part of the leader set L for the next cascade. Concretely, update the Leader Set as $L = N \cup L\backslash l \ \forall l \subseteq L$ and having an assigned follower.*

*5. Update the Follower Set as F = F\N*

*6. Repeat steps 4 and 5 until F is empty.*

*7. For every node $l \subseteq L$ (which now has only those nodes which have no followers associated with it), assign it a community membership possessed by the majority of its neighbors with ties broken randomly.*

We demonstrate the credibility of the above algorithm for community detection in the Technical Details Section. Next, we speak about the computational time complexity of the algorithm and then apply the algorithm to perform community analysis of synthetic and real-world complex networks.

## IV. COMPUTATIONAL COMPLEXITY ANALYSIS

In this section, we discuss the computational complexity of our algorithm by postulating that -

*Given a network graph G (V, E), where V and E refer to the vertex and edge set of G respectively, the proposed Cascaded Stackelberg Community Detection (CASCODE) Algorithm will detect the communities in $O(|V||E|)$ time.*

**Proof** - The computational requirements of our algorithm are dominated by the leader detection step. The leader detection step comprises of 2 stages, viz. betweenness centrality computation and leader assignment. For a network G (V, E), the asymptotic complexity to calculate betweenness scales as $O(|V||E|)$, which approximates to $O(|E|^2)$ for sparse networks [1]. This can be achieved on undirected networks using Brandes' algorithm. Leader detection is then performed by comparing the betweenness centralities between neighboring nodes. Since this involves at most 2 computations per edge (one for each node on either side of the edge), this leader assignment step is a process of complexity $O(|E|)$. Since this is less than that required to compute the betweenness centrality, the entire leader detection process completes in $O(|V||E|)$ time. The further community assignments steps occur in linear viz. $O(E)$ for every block in the Stackelberg Cascade. Thus, the worst-case computational complexity of our CASCODE algorithm is $O(|V||E|)$.

## V. RESULTS

Community Detection algorithms provide powerful tools to diagnose the characteristics of local structures in networks. However, to establish inferences based on the predicted communities, it is essential to analyze the accuracy of our algorithms. At the same time, it is essential to address the computational efficiency of the algorithms. An essential hypothesis which we realize in Community Detection is that the community structure is uniquely encoded in the network's wiring diagram. Yet, we observe that different algorithms uncover different partitions of the network. The cause of this is objective on which the algorithms optimize upon, such as modularity. Thus, in the literature of Network Science, we have 2 benchmarks, viz. Girvan-Newman (GN) and Lancichinetti-Fortunato-Radicchi (LFR) benchmarks, which are networks with predefined community structures which are used to test the accuracy of community finding algorithm.

The GN algorithm generates a random graph in which the degree distribution is essentially uniform and all communities have identical size. However, in most real-world networks have fat-tailed distribution in degree as well as community size. Hence, algorithms which perform well on GN benchmark may not perform well on the latter. The LFR benchmark alleviates this discrepancy by building reference networks in which nodes are planted such that both node degrees and community sizes follow power laws.

To compare the predicted communities with those planted in the benchmark, we use Normalized Mutual Information (NMI) as the performance metric. This metric is computed by exploiting individual community probability distributions with the joint probability distribution of both the communities. The value of NMI is unity for a perfectly accurate community detection on the benchmark and is zero if the ground truth and the community prediction is statistically independent. The algorithmic time complexity of our proposed algorithm is compared with the traditional algorithms in Table 1 below.

TABLE I: TIME COMPLEXITY OF COMMUNITY DETECTION ALGORITHMS

| Sr. No. | Algorithm | Optimization Objective | Time Complexity |
|---|---|---|---|
| 1 | CASCODE (Proposed) | Cascaded Stackelberg | $O(|E|^2)$ |
| 2 | Louvain | Modularity Optimization | $O(|E|)$ |
| 3 | Ravasz | Hierarchical Agglomerative | $O(|V|^2)$ |
| 4 | Girvan-Newman | Hierarchical Divisive | $O(|V|^2)$ |
| 5 | Greedy Modularity | Modularity Optimization | $O(|V|(\log|V|)^2)$ |
| 6 | Infomap | Flow Optimization | $O(|V|\log|V|)$ |
| 7 | Clique Percolation | Overlapping Communities | $O(e^{|V|})$ |
| 8 | Link Clustering | Hierarchical Agglomerative | $O(|V|^2)$ |

Thus, our proposed algorithm has lower time complexity than most of its counterparts, except Louvain Algorithm, for sparse networks, where $|E| << |V|$. We now assess the performance of our algorithm on a real network (Fig. 1) and some synthetic networks (Fig. 2-4).

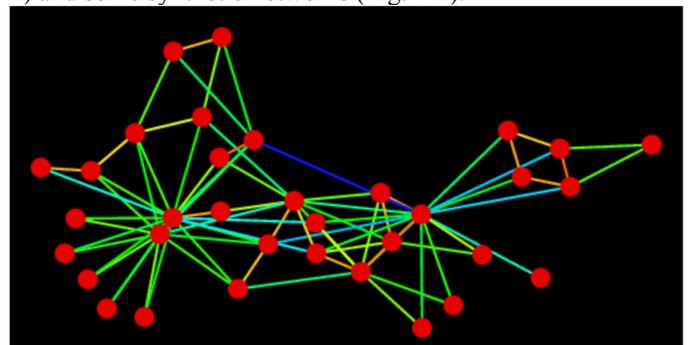

Figure 1: Zachary's Karate Club Social Network

The absolute modularity difference resulting from the use of our proposed CASCODE community detection algorithm and the Newman benchmark algorithm is tabulated in Table II. As we can see in Fig. 1-4, none of the networks possess the special structures we considered (viz. clique with minimal total internal connectivity) while building our algorithm. However, even then, in all the cases, the performance of our algorithm is similar to that of the Greedy Agglomerative Newman Algorithm. As mentioned before, our algorithm worked perfect when dealing with real evolving communities which are formed naturally by bottom up merger of nodes through their preferences.

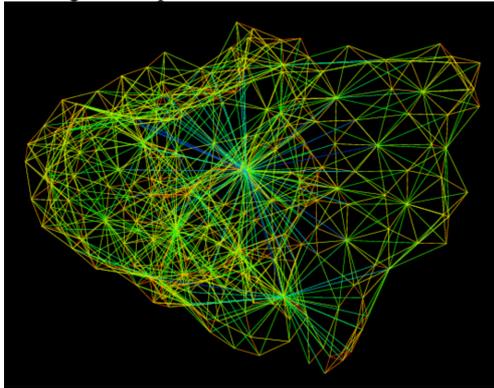

Figure 2. Synthetic Network 1

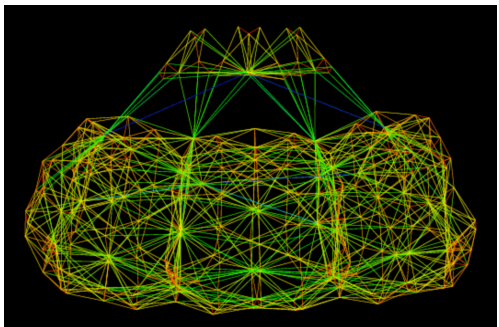

Figure 3. Synthetic Network 2

TABLE II: MODULARITY PERFORMANCE ANALYSIS OF THE CASCODE ALGORITHM ON SYNTHETIC NETWORKS

| Sr. No. | Network | | Modularity$_{CASCODE}$ − Modularity$_{NEWMAN}$ | |
|---|---|---|
| 1 | Fig. 1. | 0.03 |
| 2 | Fig. 2. | 0.02 |
| 3 | Fig. 3. | 0.04 |
| 4 | Fig. 4. | 0.05 |

We then performed a performance comparison of our proposed algorithm with the Greedy Newman algorithm in terms of NMI on Girvan-Newman Benchmark Networks. We preferred to use this benchmark to test the robustness of our algorithm to networks which do not have a defined modular structure. In every experiment, we state the absolute difference between the results of our algorithm and the Greedy Newman algorithm to demonstrate our performance with respect to the latter algorithms, even for networks lacking succinct community structure. We preferred to use the Absolute Difference as a performance assessment heuristic so as to compare the output of our algorithm to benchmark techniques, since the networks are complex for visual inferences. For every experiment, we build a GN benchmark network with a specific number of communities containing a specific number of nodes per community. The results of this experiment are tabulated below in Table III.

As seen from the results above, our results in detecting communities are in comparison with those of benchmark algorithms such as the Newman algorithm considered in our test. However, as against Newman algorithm, which scales with a function of the number of vertices in the network, our proposed algorithm scales with the number of Edges in the network giving it a computational speed advantage in case of very-large sparse networks.

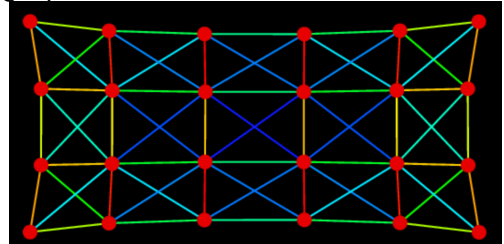

Figure 4. Synthetic Network 3

TABLE III: PERFORMANCE ANALYSIS OF THE CASCODE ALGORITHM ON SYNTHETIC NETWORKS BASED ON GN BENCHMARKS

| Sr. No. | # Nodes per Community | # Communities | | NMI$_{CASCODE}$ − NMI$_{NEWMAN}$ | |
|---|---|---|---|
| 1 | 10 | 5 | 0.04 |
| 2 | 20 | 30 | 0.03 |
| 3 | 10 | 20 | 0.02 |
| 4 | 6 | 15 | 0.02 |

We now prove the correctness of the hypothesis we used to build our CASCODE Framework.

## VI. TECHNICAL DETAILS

In this section, we give a firm footing to the hypothesis we postulated to identify leaders and their followers succinctly with an example. Concretely, we prove as to how the use of betweenness centrality correctly identifies leaders (or co-leaders) within a particular community (or a sub-community) without any false positives. We also then prove the correctness of our community assignment step as the community label propagates from the leaders (or co-leaders) to their respective followers.

Recall that we defined a follower at any stage of our Stackelberg Game as a node which has total internal connectivity with respect to the local community hosted by its immediate leader (often co-leader). At the same time, we defined a leader to have the highest betweenness centrality amongst its neighbors. To prove the credibility of our leader detection algorithm, consider the segregation of a toy network into two disjoint sets as shown in the figure. Here l refers to the leader of the community which hosts f. In the example shown below, the community under analysis is a clique on which we will base our arguments. However, as we proceed, it will be clear that our logic indeed is extendable to generic networks. Clearly, all nodes in the community are linked with each other and hence their betweenness is the base value. The node f can only reach u2 through l. Thus, it has a higher betweenness centrality than f and similar other neighbor nodes which do not have branches emanating from them, unlike the node v. This is verified from the fact that indeed f is a follower. Had it had a branch connected to it, it would be a leader as it would have a betweenness centrality proportional to the number of nodes connected in its branch. Clearly, v and l, have the same betweenness centrality, highest amongst their neighbors in the community and hence are qualified as leaders by our algorithm. Correspondingly, nodes u2 and the node connecting v to u1 will act as leaders to the respective cliques associated with them.

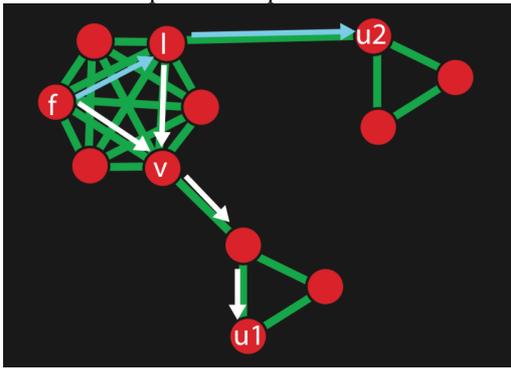

Figure 5. Network with Proposed Community Structure [1]

If they had a different number of nodes connected to these branches, say l having higher, l would have higher betweenness centrality and would go first in the community assignment step. In this case, it will assign the community label associated with it to all the nodes in the clique except v. Consequently, v will have no followers to be assigned, as they all would be associated with l. Thus then, v would lose its membership and be assigned to a community, a majority of its neighbors are connected to, in turn acting as a follower of l, and possessing its membership. Consequently, all the nodes in the other 2 cliques get their respective membership as assigned to and by the leaders hosting them. Thus, the proposed model correctly detects the community structure in the network. To summarize, this is attributed to 2 main ideas -

*1. Assigning leaders using betweenness centrality heuristic.*
*2. Assigning the followers (leader's neighbors) to follow the leader in terms of possessing the same community membership as the leader.*
*3. Assigning leaders or (co-leaders) with no followers left to be assigned to act as a follower of the leader (or co-leader) possessing highest control over its neighbors.*

Thus, the proposed algorithm will detect all of the communities exactly for networks possessing community cliques with each clique having at least one node with total internal connectivity. However, as is seen in the results, the algorithm extends naturally to arbitrary networks and produces relevant community mappings. Also, from the above approach, it is clear that we are indeed building the communities bottom-up by considering individual node preferences without optimizing on a central objective, thus exploiting the structures exhibited by evolving dynamic communities.

## VII. CONCLUSION

This article proposes a bottom-up algorithm for community detection using the Stackelberg Duopoly Model. The intuition behind the modeling the community detection problem as a Game Theoretic one is that communities evolve dynamically through node preferences and not through an imposition by central governing authority, as against ideas that influence techniques of global heuristic optimization. The proposed algorithm is a non-parametric one, and hence, it is not required to know apriori the number of communities that a network is going to contain. This information is certainly difficult to estimate in large complex dynamic networks. Thus, the CASCODE algorithm is able to learn the communities in the network, unlike spectral clustering. The bottom-up community resolution approach which forms a base of our algorithm makes it effective in detecting small-scale community structures in dense networks, where conventional techniques such as spectral clustering fail. This advantage appears because our algorithm exploits internal community characteristics as opposed to external properties used by traditional spectral clustering.

Because community detection algorithms have a wide range of applications as introduced before, we would like to survey the theoretical guarantees of the algorithm to a large set of communities, viz. a superset of the communities we considered with primacy. In general, networks are not composed of mutually exhaustive cliques, but rather dense graphs formed through preferential attachment or random growth. We have already proved that the proposed algorithm works accurately in identifying communities in networks formed by mutually exhaustive cliques connected through various leaders. However, we would like to investigate a lower bound for the edge density threshold for which a subgraph can be approximated as a clique (which has edge density 1) so that the proposed algorithm works as accurately in detecting communities. We also observe that our algorithm gives good results even in the presence of a large number of community structures. To check the robustness of our algorithm, all the cases we considered in our experimental setup were networks with ill-defined community structures. However, the results we obtained using our framework of natural community resolution, were similar to comparison with those obtained through heuristic optimization techniques. Thus, through our results, we believe that such theoretical guarantees are indeed possible for our algorithm, and will continue to be an interesting abstract research problem. Since our framework doesn't revolve around optimization of a central objective, capturing the mechanism of evolution of a community, one would also like to investigate on how to club communities resolved by our network to predict fusion in sub-community structures to form bigger communities.